# The Roles of Culture in Online User Reviews: An Empirical Investigation

Poompak Kusawat & Surat Teerakapibal

Electronic word-of-mouth (eWOM) is a prominent source of information that significantly influences consumer purchase decisions. Recent literature has extensively explored the impact of eWOM on consumers-generated reviews and purchase decisions. However, few studies have analyzed the role of culture on eWOM. We use a novel dataset of Airbnb eWOM messages in order to empirically extend the findings by Banerjee and Chai (2019). We find that the sentiment of individualistic customers is worse than that of their collectivistic counterparts when both groups experience the same level of negative disconfirmations. Furthermore, guests from a relatively more distant culture rely less on heuristics. In particular, quality signals, such as the "superhost" status, are more influential to consumers from a less distant cultural background.

Keywords: eWOM; culture; cultural distance; user-generated content; online user ratings; quality signaling; expectation-disconfirmation theory

**Introduction**

In many countries, the tourism industries contribute to their sustainable economic growth (Holzner, 2011; Kim & Chen, 2006; Teerakapibal, 2016). According to the World Tourism Organization, the direct contribution of travel and tourism to GDP was $2,750.7 billion in 2018 and was forecasted to rise to $2,849.2 billion in 2019 (Stainton, 2020). Among many significant trends in the tourism industry is the participation in peer-to-peer platforms, which connects service providers directly with their customers. One of the best-known examples of this type of platforms is Airbnb. Founded in San Francisco in 2008 as a platform where hosts list their spare rooms or residences to potential guests, Airbnb offered 7 million unique places to stay in more than 100,000 cities and 191 countries and regions in 2019 (Airbnb, 2019). In the same year, the company with more than 2 million renters per night and over 500 million users was valued at $31 billion (Airbnb, 2019), the amount that surpassed the valuation of

major hotel chains despite the company scarcely owning any considerable assets (Sigala, 2017).

Literature has documented the impacts of word-of-mouth on consumer purchase decisions and overall satisfaction (for examples, Anderson, 1998; Chu & Choi, 2011). In particular, informal communications among peers have a stronger influence when consumers evaluate alternatives in choice sets than do formal communications from the firms (Dichter, 1966; Fornell & Bookstein, 1982; Singh, 1988; Westbrook, 1987). Interestingly, online word-of-mouth is no exception. Turton (2015) suggests that user ratings and reviews are an important source of information for online consumers. Specifically, 70% of online consumers rate peer recommendations and reviews above professionally written content. The importance is even more pronounced in travel-related services because such services are often experience-based – the type of product characterized by uncertain quality (Banerjee & Chai, 2019). Furthermore, the global adoption of the internet and mobile usage encourages consumers across geographical regions to participate in and utilize online user reviews (Harris & Prideaux, 2017). The consequence is twofold: huge variation in cultural backgrounds of consumers who write and/or use reviews and a large number of user-generated reviews.

Variation in cultural backgrounds of consumers who write and/or use reviews poses challenges for marketers. In particular, Kozinets (2016) suggests that research on online word-of-mouth incorporates different cultural realities into consumer-generated reviews. Unsurprisingly, a stream of research on online word-of-mouth has emerged with the objectives of investigating how consumer culture impacts the effectiveness and participation of online reviews (for examples, Lam et al., 2009, Lin & Kalwani, 2018). However, most research on culture and online word-of-mouth merely documents the significance of the effect rather than investigating the mechanism behind it.

Unlike prior literature, Banerjee & Chai (2019) employ expectation-disconfirmation theory and prospect theory to explain a negative relationship between individualism and online review rating. Expectation-disconfirmation theory states that a customer's satisfaction is a result of the comparison between perceived performance and the customer's pre-purchase expectation of the product. On the other hand, prospect theory suggests that people evaluate their experiences in terms of gains and losses with respect to the reference point that is the expectation of product performance. These two theories therefore suggest that consumers are more likely to engage in writing negative reviews when they experience negative disconfirmations. Banerjee & Chai (2019) also posit that individualistic consumers are more likely to write more negatively when product experience is worse than what is expected. The authors conjecture that this is due to the fact that individualists' preferences vary more extensively around the communal one than those of collectivists. We contribute to the literature by demonstrating that even for the product with the same level of negative disconfirmation, individualistic customers' sentiments are worse than those of collectivistic consumers. More specifically, our empirical results show that individualism negatively moderates the relationship between negative disconfirmations and review sentiments. Disconfirmation of expectations rating reflects the difference between customers' expectations prior to their actual experiences and what their actual experiences – a measure unique to Airbnb. In this particular analysis, we utilize the Valence Aware Dictionary for Sentiment Reasoning (VADER) tool on user-generated comments to directly obtain customers' sentiments without relying on the inference from the overall rating of the product as in extant research. In addition, a hierarchical model framework is used to account for the multilevel nature of the data comprising individual-level and host-level variables.

Another consequence of growing global interests of eWOM is the huge number of user-generated reviews available to potential customers. Although information assists consumers in making well-informed decisions, "too much" information has long been argued to result in dysfunctional consequences (See Jacoby, 1984). Subsequently, prospective customers may lack sufficient cognitive capabilities to process all the data, ultimately leading to inability to perceive the true quality of service before making purchase decisions. In this scenario, Rao et al. (1999) posit that consumers will rely on signals to form expectations about the true quality of the product. This behavior is evident especially when the signal is costly (Kardes et al., 2004).

In the context of Airbnb, such a signal is whether the host is a "superhost." To become a "superhost," an owner shall meet the following arduous criteria: (1) host a minimum of 10 stays in a year, (2) respond to a guest quickly and maintain a 90% response rate or higher, (3) have at least 80% 5-star reviews, and (4) honor confirmed reservation. Hence, the "superhost" status is highly valued by consumers seeking high quality hosts. In particular, extant studies on Airbnb document the "superhost" status as a signal of quality, and its positive impacts on consumer outcomes (Liang et al., 2017; Ma et al., 2017; Wang & Nicolau, 2017; Xie & Mao, 2017). However, as global consumers consist of individuals from different cultural backgrounds it remains unclear whether the effectiveness of quality cues remains consistent across distinct cultural segments. We thus extend the literature by investigating the moderating effect of cultural distance on the relationship between the possession of the "superhost" status and customer's review sentiments. A multi-level modeling framework is used to estimate the impacts of Hofstede's cultural distance scores on review sentiments. We find that tourists from more distant cultures (i.e., tourists from the countries whose cultural backgrounds are dissimilar from those of the host country) rely less on quality cues, such as the "superhost" status, than those who travel from a culturally similar origins (i.e., tourists from the countries whose cultural backgrounds are similar to those of the host country). This finding indicates that businesses should allocate resources to the creation of quality cues only if they target customers who share their cultural profiles.

In sum, this paper aims to shed light on the effects of culture on online consumer-generated reviews and purchase decisions by answering two main research questions:

1. Does individualism negatively moderate the detrimental impact of negative disconfirmation of expectations on the sentiment of online reviews?
2. Does cultural distance attenuate the effectiveness of quality cues in promoting positive reviews?

We begin with a thorough literature review to develop related hypotheses. Then, we explain the data and methodologies used for analysis. Finally, we present the results and discuss conclusions, implications, limitations and future research directions.

**Literature review**

*Electronic Word-of-Mouth (eWOM)*

Word-of-mouth (WOM) is defined as "the exchanging of marketing information between consumers in such a way that it plays a fundamental role in shaping their behavior and in changing attitudes toward products and services (Katz & Lazarsfeld, 1966)." These exchanges allow consumers to access organic information related to the product or service, as opposed to formal advertising, which are predominantly controlled by the companies such as through television, radio, print advertisements—for example. Existing research indicates that consumers perceive WOM to be a much more credible source of information than they do traditional channels (Cheung & Thadani, 2012). This is because users in general would trust other consumers more than they do sellers (Nieto et al., 2014), since consumers are viewed as having no direct incentives to endorse the product or services. This independence makes WOM a more credible source of information (Arndt, 1967; Lee & Youn, 2009). Moreover, consumers are often only willing to disseminate information through WOM, when the source is believed to be trustworthy, since they generally want to present themselves in a positive way (Berger, 2014). Advocating false facts about a product in WOM could be perceived negatively. As a result, consumers are more likely to endorse a product or brand only when they trust its authenticity. To conclude, WOM is regarded as one of the most valuable sources of information about products and services—especially in the field of tourism and hospitality where intangible products are difficult to evaluate prior to consumption (Huete-Alcocer, 2017).

      Since the advent of online platforms, a new form of communication has become a central focus rapidly. Electronic word-of-mouth communication (eWOM) is defined by Goldsmith (2006) as "word-of-mouth communication on the Internet, which can be diffused by many internet applications such as online forums, electronic bulletin board systems, blogs, review sites, and social networking sites". A prime example of eWOM is posting reviews on online platforms (Aksoy et al., 2018; Babic Rosario et al., 2016; Fong & Burton, 2008; Huang et al., 2014). Like WOM, eWOM is recognized as an important source of products or service information that influences consumer behavior (Brown & Reingen, 1987; McFadden & Train, 1996; Messner, 2020; Zehrer et al., 2011). While many consider eWOM to be the electronic version of traditional WOM, certain differences between the two concepts are distinct. First of all, eWOM is a more powerful communication channel because it can be accessed anywhere through the internet (Bakos & Dellarocas, 2011; Duan et al., 2008). Electronic word-of-mouth is also more balanced and unbiased because various opinions are presented simultaneously on the same website (Lee et al., 2008; Senecal & Nantel, 2004). Given these unique characteristics of eWOM, simply relying on existing literature on traditional WOM would probably be insufficient to fully understand the mechanism of eWOM. eWOM research concentrates on the valence and volume of eWOM. Valence is the

direction of the eWOM communication (i.e., how positive or negative a rating is) while volume is the amount of e-WOM communication (i.e., the number of ratings) (Babić Rosario et al., 2016; Floyd et al., 2014; Purnawirawan et al., 2015; You et al., 2015). Extant literature indicates that eWOM valence is a consequence of customers' satisfaction (Boulding et al., 1993; Ho et al., 2017) and, in turn, affects consumer purchase decisions—with consumers especially paying more attention to negative information than to positive information (Cheung & Thadani, 2012).

*Quality Signaling*

Information asymmetry is when parties in a transaction have different amounts of information (Akerlof, 1978; Mishra et al., 1998). It occurs when the actual quality of a product is not readily observable because of its complex and experiential nature, or when companies do not share all product-related information with their consumers (Nelson, 1970; Tellis & Wernerfelt, 1987). In this situation, consumers' perception of product quality is impaired, increasing risk associated with their purchase decisions. The main purpose of quality signaling is to convey credible information about unobservable product quality to the consumer (Rao et al., 1999). Advertising expenditure is one of the most frequently cited quality signaling in the marketing literature. Providers of high-quality products are more incentivized to advertise than providers of low-quality products (Kirmani & Wright, 1989; Nelson, 1970) because if a large sum of money were spent on advertising and consumers realized that the true product quality was low, then they would not repeat their purchase and firms would not be able to recover costs of advertising (Milgrom & Roberts, 1986). Thus, the creditability of a signal is an important consideration—and costly signals are reliable whereas while cheap ones are less reliable as they are easily imitated.

*Culture*

Guiso et al. (2006) define culture as "those customary beliefs and values that ethnic, religious, and social groups transmit fairly unchanged from generation to generation." Extant research posits that the difference in cultural value systems causes consumer behavior to vary (Chu & Choi, 2011; Fong & Burton, 2008; Peter & Olson, 1998; Pfeil et al., 2006)—including the influence of eWOM (for examples, Christodoulides et al., 2012; Lin & Kalwani, 2018; Kim et al., 2018). In order to measure cultural differences, Hofstede's cultural dimension theory is commonly employed (Hofstede, 2001). Among various dimensions of cultures, one of the most widely used dimension is individualism-collectivism, which is related to the integration of individuals into primary groups. While individualism is defined as the preference that puts a higher degree of weight on an individual's own preference over the collective preference of a social group, collectivism is defined as the preference that puts a higher degree of weight on the collective preference of a social group over an individual's own preference. Marketing literature suggests that invidualism-collectivism dimension is a relevant determinant of consumers' behaviors. For example, Yang et al. (2019) propose a framework explaining how individualistic and collectivistic consumers are differentially motivated to pursue loyalty program rewards. Specifically, to-date (to-go) progress feedback is more suitable for individualistic (collectivistic) customers (Yang et al., 2021). Further, Yu et al. (2021) suggest that the effectiveness of managerial response to negative online reviews vary depending on consumers' cultural backgrounds.

**Hypotheses**

*Culture and Online User Reviews*

Oliver's (1977, 1980)'s expectation-disconfirmation theory is one of the most influential theoretical frameworks to explain consumer satisfaction. According to the theory, satisfaction is referred to as an affective state representing an emotional reaction to a product or service. Satisfaction is a function of the difference between pre-purchased expectation and post-purchase perceived performance of a product or service. Specifically, prior to a purchase, a customer forms a certain expectation of his/her probable experience. This expectation then serves as a comparative reference. After experiencing the product's actual performance, a customer then perceives the discrepancy between his/her expectation and the actual performance. This discrepancy could result in satisfaction or dissatisfaction. That is, if the actual performance of a product outperforms its expectation (positive disconfirmation), post-purchase satisfaction will result. If the actual performance of a product falls short of expectation (negative disconfirmation), the consumer will be dissatisfied.

In fact, expectation-disconfirmation theory has also been widely applied in e-commerce (Ahn et al., 2016; Ho et al., 2017; McKinney et al., 2002; Moe et al., 2011; Naragajavana et al., 2017; Qazi et al., 2017). More specifically, Ho et al. (2017) show that disconfirmation of expectations significantly influnces post-purchase eWOM ratings. Other prominent determinants of online review sentiment are equity (see Szymanski & Henard, 2001) and pricing structure (see Wolkenfelt & Situmeang, 2020). Relevant studies, including Banerjee & Chai (2019), further argues that the sentiments of online reviews vary with individuals of distinct cultural values (e.g., individualism vs. collectivism).

In collectivistic cultures, people exhibit more tolerance of the inconsistency between values and behavior and are more flexible to adjust their behavior to social norms (Kashima et al., 1992). Consequently, personal values are deemed less relevant to how one should behave. In contrast, individualistic agents base their behavior more readily on personal values; therefore, they are more active in self-expression than collectivistic individuals (Hofstede 2001; House et al. 2004). Findings from past studies support this notion. For instance, Kim & Sherman (2007) find that participants from individualistic cultures exhibit a stronger emphasis on expressing oneself than those from collectivistic cultures when making decisions. Furthermore, Kim et al. (1994) find that members of collectivistic cultures are more concerned with avoiding hurting others and imposing on others than are members of individualistic cultures.

In the context of online reviews, customers who have experienced negative disconfirmation would reflect their dissatisfactions through their written reviews. However, the negative sentiment of the reviews should be exacerbated for individualistic customers who are more self-expressive than collectivistic customers. Thus, we hypothesize that:

H1: The negative impact of disconfirmation on review sentiment is more salient for individualistic consumers.

*[Figure 1 near here]*

*Quality Signaling and Online User Reviews*

Along the same line, a "superhost" status in the context of Airbnb can be considered a costly signal of quality. According to the Airbnb website, the owners must satisfy rigorous conditions to be granted a "superhost" status. Therefore, to obtain and keep this badge, owners must devote valuable efforts to their listings. This clearly indicates high standards relative to "non-superhosts", producing a strong and positive signal to guests in terms of the quality provided by a "superhost" and ultimately resulting in more bookings.

This argument is consistent with the findings from previous research. For example, Xie & Mao (2017) document that host quality attributes, including being a "superhost", significantly influence listing performance. Similarly, Liang et al. (2017) show that Airbnb guests are willing to pay more to a "superhost" than to a regular host and that a "superhost" is more likely to receive reviews. Furthermore, Wang & Nicolau (2017) find a statistically significant and positive effect of the "superhost" status on listing prices. Finally, Ma et al.

(2017) find that a "superhost" possesses a significantly more extensive host profile. Based on this line of argument, this study hypothesizes that:

H2: Review sentiments of accommodations with "superhost" badges are more positive than those of accommodations without "superhost" badges.

*Culture and Quality Signaling*

Cultural distance is a concept that measures the extent to which ones' cultures of origin are different from or similar to the cultures of others (Shenkar, 2012). The construct of cultural distance has provided a tangible and quantitative measure for cultural differences by simplifying the complexities of culture (Moon & Park, 2011). One of the pioneering efforts to measure culture in a quantitative manner is the work by Hofstede. By surveying over 116,000 workers employed in a multinational company located in 72 countries between 1967 and 1973, Hofstede proposes four cultural dimensions: power distance, individualism, masculinity, and uncertainty avoidance. The fifth and sixth dimension, long-term orientation and indulgence, are later added to the framework. Later, Kogut & Singh (1988) develop cultural distance composite index by utilizing the differences in scores on Hofstede's dimensions. The vast majority of cultural distance studies follow this approach in operationalizing and measuring cultural distance.

Intuitively, substantial cultural distance between home and host countries could lead to emotional discomfort (Ye et al., 2013) because of higher risks and uncertainties. Crotts (2004) studies the overseas travel behavior of outbound travelers from the United States to 26 different countries. The results show that cultural distance is positively correlated with risk-reducing behaviors such as larger travel groups, less frequent travels, shorter trips, and the lower number of destinations in the itinerary. One possible explanation is that high cultural distance can evoke the possibility of experiencing difficulties in communicating with foreigners, cultural misunderstanding, and inability to adjust to a foreign way of life, thereby resulting in high cultural risk (Bi & Lehto, 2018).

Similarly, in the Airbnb context, when consumers perceive the traveling destination to have higher cultural distance, they will perceive higher risk perception. To reduce potential risks and uncertainties abroad, consumers generally conduct intense information search (Money & Crotts 2003; Cho & Lee, 2006). As a result, they are more willing to thoroughly process information on the accommodations and less likely to rely on heuristics, such as "superhost" status. Therefore, this study hypothesizes that:

H3: The positive effect of being a "superhost" on review sentiments is attenuated when the cultural distance between the consumer's home country and the country where accommodation is located is high.

*[Figure 2 near here]*

**Data and Measures**

We retrieve the data for this study from three different sources: insideairbnb.com, airbnb.com (the official Airbnb website), and Hofstede's cultural values.

Insideairbnb.com is an independent, non-commercial set of data that is directly scraped and archived from publicly available information on the Airbnb website. Currently, the database provides listing data and review data from numerous cities all across the globe. The sampled Airbnb listings used in this study are from Venice, Italy, during September 2019. Venice is chosen owing to a wide variety of its visitors' cultural backgrounds and the number of observations available.

Although a list of reviewers and their reviews on an Airbnb listing are readily retrievable from insideairbnb.com, the country of origin of each reviewer is not available. Rather, this information is publicly available on the user profile page on the Airbnb website.

Thus, we create a customized web scraping tool. For each reviewer, the tool directly accesses his/her profile page on airbnb.com and collects the information about his/her home country.

To capture individualism and other cultural values, we source Hofstede's index scores from geerthofstede.com, a website created and maintained by the Holfstedes themselves. Similarly, a composite index of cultural distance is calculated with Hofstede's dimensions from geerthofstede.com.

The final data set contains 233,446 customers from 6,611 Airbnb accommodations.

*Review Sentiment*

The review texts are retrieved from insideairbnb.com. From these texts, we classify them as either a positive or a negative review using VADER (Valence Aware Dictionary for Sentiment Reasoning), a lexicon and rules-based sentiment analysis tool that is specifically attuned to sentiments expressed in social media (Hutto & Gilbert, 2014). Lexicon based sentiment analysis embodies a list of lexical features, e.g., words, that are generally labeled according to their semantic orientation as either positive or negative. In the case of VADER, the intensity of positive or negative words is also taken into account. Moreover, VADER is intelligent enough to recognize emoticon, capitalization, punctuation, and words that reverse the polarity (such as 'not', but', etc.) and change the intensity and polarity accordingly. VADER produces four sentiment metrics from these word ratings. The first three, i.e., positive, neutral and negative, represent the proportion of the text that falls into those categories. The final metric, compound score, is the sum of all of the lexicon ratings which have been normalized to range between -1 and 1. Using the compound score, we classify the review as either a positive or a negative sentiment with classification thresholds set at above +0.05 or below -0.05, respectively. These thresholds are utilized by the original authors of VADER. Their study shows that VADER's performance is on par with a sophisticated machine learning-based approach and outperforms other lexicon-based approaches (Hutto & Gilbert, 2014). VADER is specifically suitable for our online review dataset because its lexicon is especially attuned to microblog-like contexts and performs exceptionally well in the social media domain.

Since some of these reviews are written in languages other than English, we must translate these reviews into English before analyzing their sentiments. First, we identify reviews that are not in English by following an approach proposed by Lui & Baldwin (2014). They compare several language detection systems and suggest that a simple majority vote across three specific tools is the best performer. These tools include langid.py (Lui & Baldwin, 2012), an n-gram feature set selected using data from multiple sources; a multinomial naive Bayes classifier "CLD2" (McCandless, 2010), the language identifier embedded in the Chrome web browser that uses a naive Bayes classifier and script-specific tokenization strategies; and LangDetect (Nakatani, 2010), a naive Bayes classifier, using a character n-gram based representation without feature selection, with a set of normalization heuristics to improve accuracy. After we flagged non-English reviews, we then translated them using Google cloud translation API, an interface for translating an arbitrary string into any supported language by using the state-of-the-art Neural Machine Translation.

*Individualism*

This study operationalizes individualism using Hofstede's index scores. The scores range from 1 to 100 with a higher score indicating a higher degree of individualism.

*Negative Disconfirmation*

For each listing on the Airbnb website, the platform allows previous customers to rate their overall experiences with the listing. In the data from insideairbnb.com, the rating ranges between zero and ten. Moreover, the customers can rate the quality of a listing for various aspects, e.g., how clean the accommodation is, or whether the hosts are helpful. One dimension of these ratings that is relevant to our study is the accuracy rating which is uniquely available on the Airbnb platform. This rating reflects the discrepancy between the accommodation description on the website and the actual product offering. A perfect score on this rating means that expectations are met while any score less than perfect implies that expectations exceed the actual experiences. This rating is the opposite of negative disconfirmation. Put it differently, a higher accuracy rating implies a lower degree of negative disconfirmation. Therefore, we reverse-code the measure by subtracting its value from the maximum score of ten.

*Quality Signaling*

Quality signaling is an indicator denoting whether the host of a listing is a "superhost" or not. Since the owner must meet specific conditions to attain a "superhost" status, this status then represents how much the host dedicates his/her effort to his/her listing. However, it does not signal to potential customers any tangible characteristics of the listing. Consequently, the "superhost" status is an appropriate measure for quality signal. This information is readily available on the database from insideairbnb.com.

*Cultural Distance*

Kogut & Singh (1988) create a composite index of cultural distance based on cultural dimensions of Hofstede. The Hofstede measure is based on six cultural dimensions: power distance, individualism, masculinity, uncertainty avoidance, long-term orientation, and indulgence. Four of these dimensions are from surveys conducted during the late 1960s and the early 1970s, while the remaining two dimensions are added in later years. Based on these cultural values, we calculated a composite index of cultural distance by following Kogut &

Singh (1988)'s formula, shown below in Equation 1. The deviations are corrected for differences in the variance of each dimension and arithmetically averaged:

$$CD_j = \frac{\sum_{i=1}^{9}[(I_{ij}-I_{ik})^2/V_i]}{9},\qquad(1)$$

where

$CD_j$ = cultural distance between country $j$ and country $k$;

$I_{ij}$ = GLOBE score for dimension $i$ of home country $j$;

$I_{ik}$ = GLOBE score for dimension $i$ of host country $k$, Italy in this case; and

$V_i$ = variance of the GLOBE score for dimension $i$.

Detailed data summary is shown in Table 1 and Table 2.

*[Table 1 near here]*

*[Table 2 near here]*

**Models**

*Model 1*

The objective of the first model is to analyze the main effect of negative disconfirmation on review sentiment and the moderating role of individualism. Our dataset possesses a two-level nested structure. Each observation is represented by an $(i, j)$ pair, where $i = 1, 2, ..., 233,446$ customers; and $j = 1, 2, ..., 6,611$ Airbnb accommodations. At the lower level of nesting, individual customers who stay in the same accommodation share common disturbances at the accommodation level. Thus, their review sentiments correlate more closely than sentiments from customers who stay in different accommodations. In other words, customers are nested within accommodation.

We describe our model using three linked equations even though they can be substituted into a single equation:

$$y_{ij} = \beta_{0j} + \beta_{1j}x_{1ij} + \beta_{2j}z_{1ij} + \cdots + \beta_{6j}z_{5ij} + e_{ij}, \tag{2}$$

$$\beta_{0j} = \gamma_{00} + \gamma_{01}w_j + u_{0j}, \tag{3}$$

$$\beta_{1j} = \gamma_{10} + \gamma_{11}w_j + u_{1j}, \tag{4}$$

where

$y_{ij}$ = review sentiment of customer *i* at accommodation *j*;

$x_{1ij}$ = individualism value of customer *i* at accommodation *j*;

$z_{1ij} \ldots z_{5ij}$ = other Hofstede's cultural values, i.e., power distance, masculinity, uncertainty avoidance, long term orientation and indulgence of customer *i* at accommodation *j*.

$e_{ij}$ = the accommodation–customer-specific error term;

$w_j$ = negative disconfirmation of accommodation *j*; and

$[u_{0j}, u_{1j}]'$ = the matrix of the accommodation-specific error terms.

Equation 2 models the review sentiment for customer *i* in hotel *j* as a function of the accommodation effect (the random intercept term, $\beta_{0j}$), Hofstede's individualism value of customer *i* ($x_{1ij}$), the **Z** vector of five other Hofstede's cultural values of that customer, and a customer-level error term. Equation 3 models the accommodation effect (the random intercept from the previous equation) as a function of its grand mean ($\gamma_{00}$), negative disconfirmation of the accommodation ($w_j$), and an accommodation-level error term. Equation 4 models the accommodation-specific individualism effect (the random coefficient of $x_{1ij}$ from Equation 2) as a function of its grand mean ($\gamma_{10}$), negative disconfirmation of the accommodation ($w_j$), and an accommodation-level error term. Of particular interest are the $\gamma_{01}$ term, which represents the impact of negative disconfirmation on review sentiments, the $\gamma_{10}$ term, which represents the grand mean of the individualism causal impact, and the $\gamma_{11}$ terms, which represents the moderation effect of individualism on negative disconfirmation.

*Model 2*

The second model aims to estimate the effects of being a "superhost" on listing occupancy rates and whether such effects are moderated by cultural distance. Like Model 1, this model includes a two-level nested structure where customers are nested within accommodation. Thus, we represent each observation by an (i, j) pair, where i = 1, 2, ..., 233,446 customers and j = 1, 2, ..., 6,611 Airbnb accommodations. The model can be described by the following equations:

$$y_{ij} = \delta_{0j} + \delta_{1j} CD_{ij} + \varepsilon_{ij}, \tag{5}$$

$$\delta_{0j} = \eta_{00} + \eta_{01} D_j + v_{0j}, \tag{6}$$

$$\delta_{1j} = \eta_{10} + \eta_{11} D_j + v_{1j}, \tag{7}$$

where

$y_{ij}$ = review sentiment of customer *i* at accommodation *j*;

$CD_{ij}$ = cultural distance value of customer *i* at accommodation *j*;

$\varepsilon_{ij}$ = the accommodation–customer-specific error term;

$D_j$ = dummy variable indicating whether accommodation *j* has a "superhost" status;

and

$[v_{0j}, v_{1j}]'$ = the matrix of the accommodation-specific error terms.

Equation 5 models the review sentiment for customer *i* in hotel *j* as a function of the accommodation effect (the random intercept term, $\delta_{0j}$), cultural distance value of customer *i* ($CD_{ij}$), and a customer-level error term. Equation 6 models the accommodation effect (the random intercept from the previous equation) as a function of its grand mean ($\eta_{00}$), "superhost" status of the accommodation ($D_j$), and an accommodation-level error term. Equation 7 models the accommodation-specific individualism effect (the random coefficient of $CD_{ij}$ from Equation 2) as a function of its grand mean ($\eta_{10}$), "superhost" status of the accommodation ($D_j$), and an accommodation-level error term. Coefficients on which we focus are the impact of "superhost" status on review sentiment ($\eta_{01}$), and the moderating effect of cultural distance value ($\eta_{11}$).

**Results**

*Model 1*

*[Table 3 near here]*

In Table 3, we present the parameter estimates for the hierarchical model as specified by Equations 2 to 4. We first discuss the variance components first and then the main and interaction effects between individualism and negative disconfirmation that are of central interest.

In Equation 3, the accommodation effect consists of a fixed part (grand mean and the observed level of negative disconfirmation) and a random part. As evidenced by the significant estimate of the variance of $u_{0j}$ ($\sigma^2(u_{0j}) = .302$, $p < .01$), there are differences in review sentiment across accommodations, highlighting the need to incorporate the multi-level structure of the data into our model. In Equation 4, the individualism effect also consists of a fixed part (the grand mean and the observed level of negative disconfirmation) and a random part. On the contrary, we fail to conclude statistical significance for the random effect ($\sigma^2(u_{1j}) = .001$, $p > .10$). Therefore, the assumption of varying effects of individualism on review sentiment is not warranted.

On the other hand, in the fixed part of Equation 3, the coefficient of negative disconfirmation is significantly negative ($\gamma_{01} = -.518$, $p < .01$). This shows that stronger negative disconfirmation will result in more negative reviews. As shown by the interaction term between negative disconfirmation and individualism ($\gamma_{11} = .002$, $p < .01$), the negative effect of negative disconfirmation on review sentiment is more salient as individualism increases, supporting H1. This exacerbation is because individualistic consumers are more self-expressive and assertive than collectivistic consumers. Interestingly, the grand mean in Equation 4 is positive and significantly different from zero ($\gamma_{10} = .012$, $p < .01$). This means that individualistic customers are more likely to write a positive review. This is due to the fact that, in our dataset, individualistic consumers encounter accurate accommodations more often than collectivistic consumers. To prove this statement, we conduct a paired-samples t-test to compare the mean individualism score of customers in accommodations with zero negative disconfirmation (M = 72.677) and that of customers who rent accommodations with some degree of negative disconfirmation (M = 70.232). The difference is statistically significant as expected ($p < 0.01$). It is, therefore, not surprising that individualistic guests write more positive reviews than collectivistic guests.

In addition, empirical results show that guests from countries with the higher level of masculinity ($\beta_{3j} = .002$, $p < .10$) and the higher level of indulgence ($\beta_{6j} = .003$, $p < .05$) are associated with positive reviews.

*Model 2*

*[Table 4 near here]*

Table 4 shows the estimation results of Model 2. We first discuss the variance components and then the main and the interaction effects of our interests.

In Equation 5, the accommodation effect consists of a fixed part (grand mean and the observed level of cultural distance) and a random part. The significant estimate of the variance of $v_{0j}$ ($\sigma^2(v_{0j}) = .657$, $p < .01$) indicates the differences in review sentiment across accommodations. In Equation 6, the cultural distance effect consists of a fixed part (the grand mean and the observed "superhost" status) and a random part. Again, we find statistical significance for the random effect ($\sigma^2(v_{1j}) = .064$, $p < .01$). This points out the varying effects of cultural distance on reviews. In sum, these results show that multi-level modeling is appropriate for our data structure.

In the fixed part of Equation 6, we find the positive and significant effect of "superhost" status ($\eta_{01} = .331$, $p < .01$). This indicates that being a "superhost" is a signal of having superior experience to offer potential customers and hence encourages more positive review sentiment. This result supports H2. In addition, the main effect of being a "superhost" on review sentiment is negatively moderated by cultural distance ($\gamma_{11} = .002$, $p < .01$). This result is consistent with H3, suggesting that travelers from culturally distant origins have a lower tendency to rely on the use of heuristics upon choosing their accommodations. This is because traveling to a culturally distant destination induces the perception of high risks. As perceived risk increases information search, these travelers are more willing to conduct thorough information processing on the accommodations and less likely to rely on heuristics, such as the "superhost" status. Lastly, we find that travelers from countries with higher cultural distance tend to write more negative reviews.

**Discussion**

The present article contributes to the growing literature on quality signaling, eWOM, and cultural value. Previous research in the area of online reviews has rarely focused on the effects of consumer cultures on online reviews and purchase decisions. One exception is the study by Banerjee & Chai (2019), who suggests that expectation-disconfirmation theory and prospect theory are the mechanism driving the negative relationship between individualism and online review rating. The authors suggest that individualistic consumers generally write more negative reviews as their preferences vary widely around the mean in comparison to that of collectivists. We contribute to the literature by demonstrating that even for the case in which the level of disconfirmation is identical between individualists and collectivists, individualistic customers' sentiments remain worse than those of collectivists using the direct measure for disconfirmation of expectation. We argue that individualistic customers more readily act upon their thoughts and emotion more readily, while collectivistic customers exhibit more tolerance and adjust their behavior to social norms. Thus, individualistic customers would foster more self-expression and assertiveness, prompting more negative reviews when they encounter negative disconfirmations. This is essentially important for practitioners as extant literature suggests that online user ratings have a significant impact on company sales (Babić Rosario et al., 2016; Floyd et al., 2014; You et al., 2015), especially in the tourism and travel industry (Zhu & Zhang, 2010). Our findings also highlight that customers from individualistic cultures may cause greater difficulty in eWOM marketing because they have a higher tendency to write negative reviews when experiencing negative disconfirmations. These findings suggest that hosts mainly serving individualistic guests align customers' expectations with the actual product experiences and that reducing negative disconfirmation potentially mitigates the tendency to write negative reviews.

Additionally, many of the existing studies have documented a "superhost" status as a signal of quality and its positive impacts on consumer purchase decisions (Liang et al., 2017; Ma et al., 2017; Wang & Nicolau, 2017; Xie & Mao, 2017). However, the effects of quality signaling on purchase decisions across cultures have not yet been well examined. Using a publicly available data from multiple sources, this paper investigates how the effects of quality signaling on customers' choices vary across different cultural distances. Since the consumption of a culturally distant product induces the perceptions of higher risk due to a language barrier, a cultural misunderstanding, and inability to acclimate to a foreign way of life. Also, as perceived risk increases information search, they are more willing to conduct thorough information processing on the product or service. Consequently, consumers cease to rely on heuristics, thereby attenuating the effectiveness of quality cues. Our empirical results support this notion. Although Airbnb listings with the "superhost" status are associated with higher review sentiments, those with customers from highly differentiated cultural origins do not benefit as much from this particular quality cue. Our findings are relevant to practitioners. Though several studies denote the benefits of being a "superhost," our study stresses that it is not always worthwhile to exert time, money and efforts to attain the badge. Specifically, Airbnb hosts in some locations where most guests are from origins of similar cultures realize the full benefits of being a "superhost," while those whose guests are primarily from somewhat different cultural backgrounds witness limited returns on investment in "superhost" badges. Moreover, because of the relatively more positive effect of being a "superhost" on review sentiments, it is not surprising that hosts with customers from lower cultural distance origins thrive to acquire "superhost" badges. Conversely, our findings suggest that hosts with relatively higher cultural distance customers focus more on building their accommodation's tangible qualities because consumer's decision making relies less on the "superhost" badge. In extreme cases, the effect differences could be exacerbated in cities with only local tourists, as opposed to cities with only foreign tourists.

    However, this present study is not without limitations. The first limitation is that the original accuracy rating score on the Airbnb website ranges from zero to five with one decimal place, whereas the data we retrieved from insideairbnb.com ranges from zero to ten without decimal place. This is because the data on insideairbnb.com multiplies the original

value by two and rounds the decimal place. Future research may attempt to retrieve the original value and estimate a more accurate relationship. Second, this study centers on Venice for analysis. Although Venice is selected because of a variety of its visitors' cultural backgrounds, future research may opt to investigate the effects of culture for a particular segment of travelers. Last, the period of our study precedes the COVID-19 pandemic, future research may attempt to investigate whether and how the findings will change if the period of the global pandemic is taken into account.

This paper empirically draws an unexplored connection between the influences of cultures on pre-purchase decisions and on post-purchase satisfaction. While the endeavor for superior quality signaling typically attracts customers, it may end in vain without thorough consideration of their cultural backgrounds. Even after customers did purchase and experienced the product/service, ironically, the negligence of cultural aspects may again result in negative word of mouth from the customers, and the loss possibly offsets the gain. In summary, not only do our findings highlight the crucial importance of cultures in the context of Airbnb, but they remind global-minded marketers of the critical influences of cultures on the whole marketing funnel.

Table 1. Data summary.

| Variable | Mean | SD | Min | Max |
|---|---|---|---|---|
| Review sentiment | 0.98 | 0.15 | 0.00 | 1.00 |
| Individualism | 72.10 | 22.21 | 12.00 | 91.00 |
| Power distance | 48.29 | 16.59 | 11.00 | 104.00 |
| Masculinity | 57.04 | 13.06 | 5.00 | 110.00 |
| Uncertainty avoidance | 58.44 | 20.44 | 8.00 | 112.00 |
| Long-term orientation | 47.29 | 21.63 | 13.00 | 100.00 |
| Indulgence | 57.09 | 16.75 | 0.00 | 100.00 |
| Disconfirmation | 0.26 | 0.51 | 0.00 | 8.00 |
| Superhost | 0.43 | 0.49 | 0.00 | 1.00 |
| Cultural distance | 1.15 | 0.63 | 0.00 | 3.78 |

Table 2. Number of Airbnb customers by countries.

| Country | Number of customers | Country | Number of customers |
|---|---|---|---|
| Argentina | 3526 | Malta | 83 |
| Australia | 9623 | Mexico | 3517 |
| Austria | 2419 | Morocco | 161 |
| Bangladesh | 18 | Netherlands | 2671 |
| Belgium | 2161 | New Zealand | 1867 |
| Brazil | 3465 | Norway | 805 |
| Bulgaria | 434 | Pakistan | 77 |
| Canada | 13750 | Peru | 431 |
| Chile | 1346 | Philippines | 398 |
| China | 5425 | Poland | 1396 |
| Colombia | 1748 | Portugal | 839 |
| Croatia | 306 | Romania | 746 |
| Czech Republic | 349 | Russia | 4012 |
| Denmark | 935 | Serbia | 143 |
| El Salvador | 59 | Singapore | 1612 |
| Estonia | 163 | Slovakia | 334 |
| Finland | 809 | Slovenia | 154 |
| France | 26191 | South Korea | 3308 |
| Germany | 14022 | Spain | 6158 |
| Greece | 661 | Sweden | 1149 |
| Hong Kong | 1194 | Switzerland | 3746 |
| Hungary | 584 | Taiwan | 846 |
| India | 2220 | Thailand | 305 |
| Indonesia | 268 | Trinidad and Tobago | 29 |
| Ireland | 1326 | Turkey | 473 |
| Italy | 11647 | United Kingdom | 27187 |
| Japan | 1024 | United States | 63235 |
| Latvia | 255 | Uruguay | 359 |
| Lithuania | 357 | Venezuela | 33 |
| Luxembourg | 185 | Vietnam | 215 |
| Malaysia | 687 | | |

*$N$ = 233446

Table 3. Estimation results of Model 1: Effects of negative disconfirmation and culture on review sentiment.

| Variable | Coefficient | Estimate | SE |
|---|---|---|---|
| Intercept | | | |
|    Fixed effect | $\gamma_{00}$ | 3.079 *** | 0.179 |
|    Random effect | $\sigma^2(u_{0j})$ | 0.302 *** | 0.105 |
| Negative disconfirmation | $\gamma_{01}$ | -0.518 *** | 0.053 |
| Individualism | | | |
|    Fixed effect | $\gamma_{10}$ | 0.012 *** | 0.001 |
|    Random effect | $\sigma^2(u_{1j})$ | 0.001 | 0.001 |
| Negative disconfirmation x Individualism | $\gamma_{11}$ | -0.002 *** | 0.001 |
| Power distance | $\beta_{2j}$ | -0.001 | 0.001 |
| Masculinity | $\beta_{3j}$ | 0.002 * | 0.001 |
| Uncertainty avoidance | $\beta_{4j}$ | 0.000 | 0.001 |
| Long-term orientation | $\beta_{5j}$ | -0.001 | 0.001 |
| Indulgence | $\beta_{6j}$ | 0.003 ** | 0.001 |
| Fit indices | | | |
|    corrected Akaike Information Criterion (AICc) | | 1541908 | |
|    Bayesian information criterion (BIC) | | 1541939 | |

*p < .10. **p < .05. ***p < .01.

Table 4. Estimation results of Model 2: Effects of the "superhost" status and cultural distance on review sentiment.

| Variable | Coefficient | Estimate | SE |
|---|---|---|---|
| Intercept | | | |
|    Fixed effect | $\eta_{00}$ | 3.550 *** | 0.035 |
|    Random effect | $\sigma^2(v_{0j})$ | 0.657 *** | 0.076 |
| "Superhost" status | $\eta_{01}$ | 1.331 *** | 0.076 |
| Cultural distance | | | |
|    Fixed effect | $\eta_{10}$ | -0.150 *** | 0.023 |
|    Random effect | $\sigma^2(v_{1j})$ | 0.064 ** | 0.018 |
| "Superhost" status x Cultural distance | $\eta_{11}$ | -0.143 *** | 0.052 |
| Fit indices | | | |
|    corrected Akaike Information Criterion (AICc) | | 1550244 | |
|    Bayesian information criterion (BIC) | | 1550275 | |

*$p < .10$. **$p < .05$. ***$p < .01$.

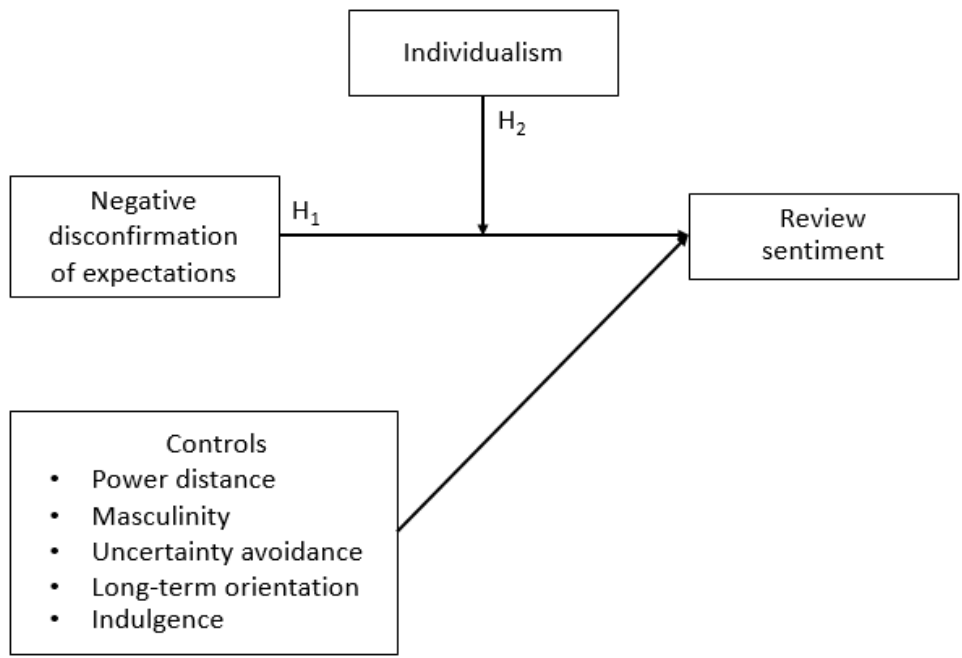

Figure 1. Conceptual framework for the relationship between negative disconfirmation and review sentiment.

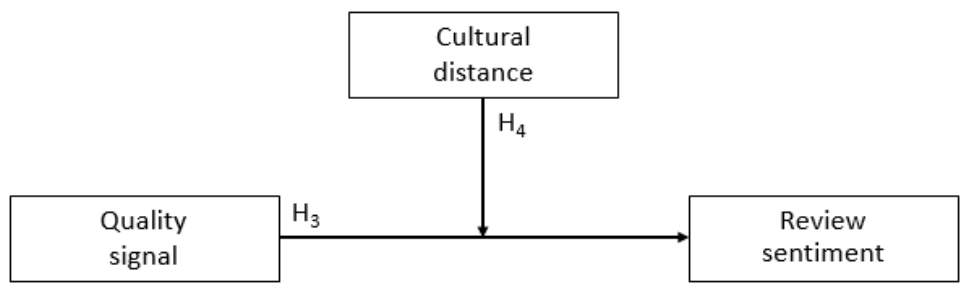

Figure 2. Conceptual framework for the moderating effect of cultural distance on the relationship between quality signal and review sentiment.

35